\documentclass[aps,prb,twocolumn,showpacs,superscriptaddress]{revtex4}
\usepackage{graphicx,amssymb,amsmath}
\usepackage{natbib}
\usepackage{bm}
\newcommand\CuFeGeO{Cu$_2$Fe$_2$Ge$_4$O$_{13}$}
\newcommand\CuScGeO{Cu$_2$Sc$_2$Ge$_4$O$_{13}$}
\newcommand\CuMGeO{Cu$_2M_2$Ge$_4$O$_{13}$}

\begin{document}
\title{Magnetic excitation in a new spin gap compound 
Cu$_2$Sc$_2$Ge$_4$O$_{13}$: Comparison 
to Cu$_2$Fe$_2$Ge$_4$O$_{13}$}
\author{T. Masuda}
\email[]{tmasuda@yokohama-cu.ac.jp}
\affiliation{International Graduate School of Arts and Sciences,
Yokohama City University, 22-2, Seto, Kanazawa-ku,
Yokohama city, Kanagawa, 236-0027, Japan}
\altaffiliation{Previous 
address: Condensed Matter Science
Division, Oak Ridge National Laboratory, Oak Ridge, TN 37831-6393,
USA} 

\author{G. J. Redhammer}
\affiliation{Department of Material Science, Division of Mineralogy, 
University of Salzburg, 
Hellbrunnerstr. 34, Salzburg A-5020, Austria}

\date{\today}

\begin{abstract}
The compound \CuScGeO\ is presented as a new member 
of the family of weakly coupled spin chain and 
dimer compounds \CuMGeO. Magnetic susceptibility, 
heat capacity, and neutron inelastic scattering 
measurements reveal that the compound has the same 
spin dimer component as \CuFeGeO. The observed narrow 
band excitation in bulk measurements is consistent 
with spin gap behavior. The energy scale of the 
weakly coupled dimers in the Sc compound is perfectly 
coincident with that in the Fe compound. 

\end{abstract}


\maketitle

\section{introduction}
In the last few decades, low-dimensional 
spin frameworks have played key roles in 
condensed matter science. Some of the 
remarkable phenomena on such spin frameworks 
include high critical temperature superconductivity 
in doped two-dimensional antiferromagnets,~\cite{Muller,Tokura}
exotic superconductivity in spin ladders,~\cite{Dagotto,SrCuO} 
and multiferroics in frustrated compounds.~\cite{Lawes05} 
A fundamental topic from the point of magnetism 
underlying these phenomena 
is the ordered or disordered in the ground state. 
In some spin frameworks, the ground state is strongly 
disordered by quantum fluctuation and is characterized 
by associated spin gap excitation. The spin liquid state, 
on the other hand, is relatively robust with respect to 
external perturbation, and thus the spin correlation 
remains short-ranged even at zero temperature. 
In conventional networks, such as chains or planes 
with gapless excitation, the ordered states are induced at finite temperatures. 

Bicomponent systems combining these distinct spin 
frameworks have recently been realized experimentally 
using R$_2$BaNiO$_5$,~\cite{Zheludev,Zheludev96a} 
\CuFeGeO,~\cite{Masuda} and Cu$_2$CdB$_2$O$_6$.~\cite{Hase} 
In the cooperative ordered state of such bicomponent systems, 
one component is strongly bound to a nonmagnetic singlet 
state, while the other maintains the usual ordered state. 
The weakly coupled compound of Cu dimers and Fe chains, 
\CuFeGeO, exhibits two spin excitations with separate 
energy scales, clearly observable by neutron scattering 
experiment.~\cite{Masuda2} In the low-energy range below 
10~meV, the excitations are well explained by effectively 
coupled Fe chains. The dimers behave as media that 
transfer exchange coupling between Fe chains, and the 
intradimer interaction can be included in effective 
interchain coupling. At higher energy, the dispersionless 
excitation observed at $\hbar \omega \sim $ 24~meV is 
qualitatively explained by the effect of Cu dimers. 
The discovery of a new member of the bicomponent 
compound family has been eagerly awaited in order 
continue systematic study of exotic types of ordering. 

\begin{figure}
\begin{center}
\includegraphics[width=7.5cm]{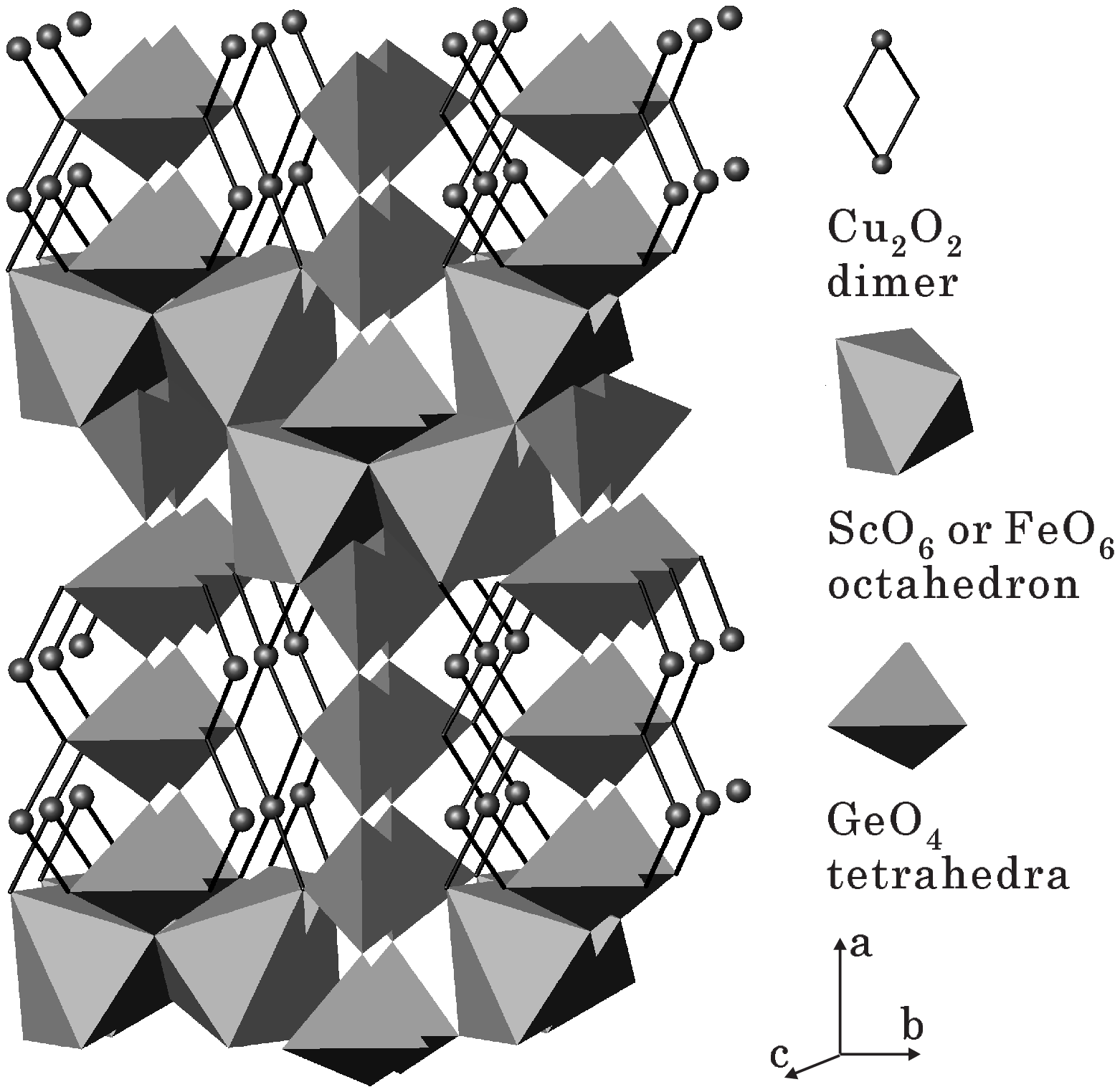}
\end{center}
\caption{(Color online) Crystal structure of 
isostructural compounds \CuScGeO\ and \CuFeGeO  } \label{fig1}
\end{figure}

One of the present authors (G.J.R.) recently 
identified a new isostructural compound \CuScGeO.~\cite{Redhammer} 
The crystal structure is monoclinic, with structural 
parameters $a = 12.336(2)$~\AA , $b=8.7034(9)$~\AA , $c = 4.8883(8)$~\AA, 
and $\beta = 95.74(2)$, and space group $P2_1/m$. 
A schematic drawing of the structure is shown in 
Fig.~\ref{fig1}. The crankshaft chains of FeO$_6$ in 
\CuFeGeO\ are replaced by non-magnetic ScO$_6$ chains, 
and the remnant Cu dimers are separated by GeO$_4$ 
tetrahedra and ScO$_6$ chains. All dimers are equivalent. 
On the basis of this crystallography, \CuScGeO is a 
candidate for a mono-component system with a single 
spin-gap framework analogous to that of \CuFeGeO. 
In the present study, \CuScGeO\ is characterized by 
a combination of bulk measurements and thermal neutron 
scattering experiment. Supplemental bulk measurements 
are also performed on the corresponding Fe compound 
in order to refine the estimated exchange constants. 
It is demonstrated that \CuScGeO\ and \CuFeGeO\ are 
ideal compounds for the systematic study of bicomponent 
systems with spin-gap and gapless frameworks. 

\section{Experimental Details}
A polycrystalline sample of \CuScGeO\ was prepared 
by a solid-state reaction route.~\cite{Redhammer} 
A high-quality sample was obtained after several 
regrind and reheat procedures. As the compound 
exhibits incongruent melting, a single crystal 
is not available at present. For the Fe compound, 
a single crystal was obtained by the floating zone 
method, and 20~mg of the crystal was carefully cut 
from the rod-shaped bulk crystal so as to ensure 
that the surface would be free of magnetic impurities. 
The crystallographic $c$ direction of these compounds 
is determined by the edge of a pair of cleavage 
planes $\{ 1~1~0\} $. The magnetic susceptibility of 
both the Sc and Fe compounds were measured under a 
1000~Oe field using a commercial SQUID magnetometer 
(MPMS, Quantum Design Co. Ltd.). Heat capacity 
measurements were performed using a commercial 
calorimeter (PPMS, Quantum Design Co. Ltd.). 
For neutron scattering experiment, 50~g of the Sc 
polycrystalline sample was placed in an Al can. 
The neutron inelastic scattering experiment was 
performed using the PONTA spectrometer at the 5G 
beamline of the JRR3 JAEA Tokai. Sollar collimation 
was employed in open - 40' - 40' - 80' configuration 
with $E_f = 13.5 $~meV, and a pyrolytic graphite 
filter was inserted after the sample to eliminate 
contamination from higher harmonics. A closed-cycle 
helium refrigerator was used to achieve low 
temperatures. Spurious peaks due to the 
Sc incoherent scattering, 
which were accidentally observed in
case of $2k_i = 3k_f$, is subtracted from 
all constant-$q$ scans. A vanadium incoherent 
scan was performed to verify the 
calculations of instrumental resolution. 

\section{Results}
\begin{figure}
\begin{center}
\includegraphics[width=8.5cm]{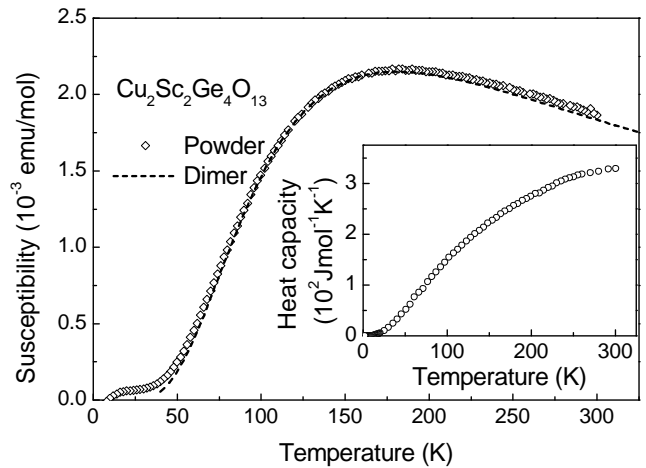}
\end{center}
\caption{
Magnetic susceptibility of \CuScGeO. 
Dotted line denotes theoretical calculation 
(see text). Inset shows heat capacity in 
zero field. } \label{fig2}
\end{figure}

The magnetic susceptibility of \CuScGeO\ 
after Curie term subtraction is shown in 
Fig.~\ref{fig2}. A broad maximum at 
170~K and a dramatic decrease at low 
temperature are observed. No phase 
transition was detected down to 2~K. 
This thermally activated behavior suggests 
the existence of a spin gap. The heat 
capacity in the zero field is shown in the inset. 
The absence of a phase transition was 
consistently confirmed in these scans.  

\begin{figure}
\begin{center}
\includegraphics[width=7.5cm]{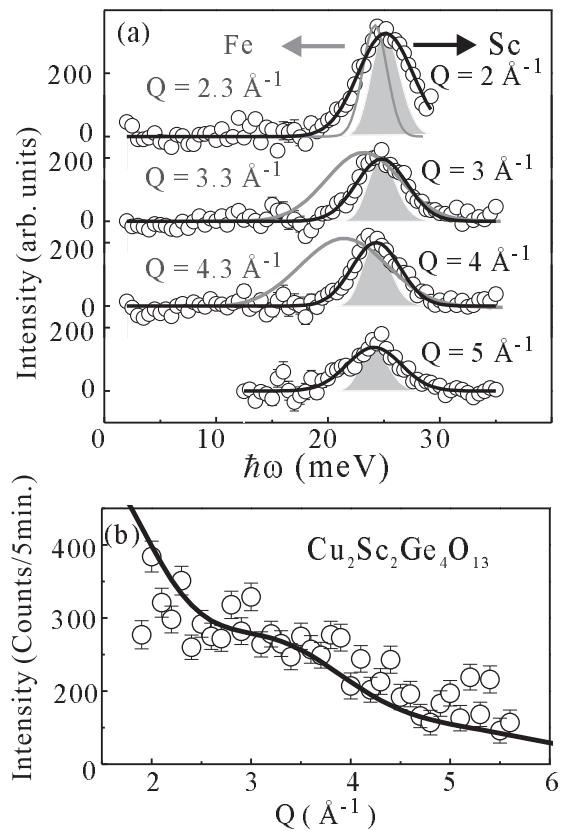}
\end{center}
\caption{(a) Constant-$q$ scans at 5~K. 
Black solid lines denote simple Gaussian 
fits, shaded areas represent instrumental 
resolution, and gray lines denote dimer 
excitation in the Fe compound from Ref.~\onlinecite{Masuda2} . 
(b) Constant-energy scan at 5~K. Solid lines 
denotes theoretical calculation (see text). } \label{fig3}
\end{figure}
A series of constant-$q$ scans at 5~K is 
shown in Fig.~\ref{fig3}(a). The scans are 
shown after linear background subtraction. 
Well-defined peaks can be observed at energy 
transfer of 25~meV over all values of $q$. 
The peak width is close to that of the 
calculated resolution function, and the 
peak energy of the present compound is 
coincidence with that of the corresponding 
Fe compound.~\cite{Masuda2} The 
constant-energy scan at $\hbar \omega = 25$~meV 
is shown in Fig.~\ref{fig3}(b) after background subtraction. 
We used the constant-$q$ 
scan at $\hbar \omega = 20$~meV as the background. 
The intensity exhibits a monotonic 
weakening with increasing $q$.
\begin{figure}
\includegraphics[width=7.5cm]{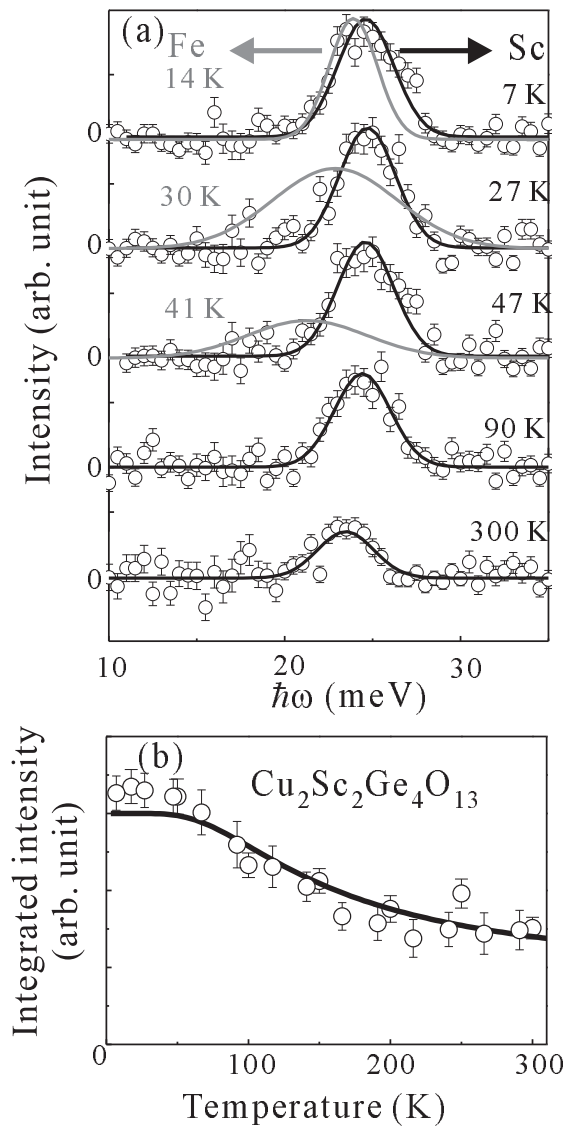}
\caption{(a) Temperature scans for 
Sc compound (open circles) and Fe 
compound (gray lines),~\cite{Masuda} 
obtained at $q = 3$~\AA$^{-1}$ and 
$q = 2.3$~\AA$^{-1}$, respectively. 
Solid lines denote Gaussian fits. (b) 
Temperature dependence of integrated 
intensity for \CuScGeO. Solid line denotes 
theoretical calculation. } \label{fig4}
\end{figure}
The temperature dependence of the 
energy scan at $q =3$~\AA$^{-1}$ is 
shown in Fig.~\ref{fig4}(a). 
Well-defined peaks with constant width 
are observable up to 300~K. The temperature 
dependence of the integrated intensity for 
the Sc compound is summarized in Fig.~\ref{fig4}(b). 

\begin{figure}
\begin{center}
\includegraphics[width=8.5cm]{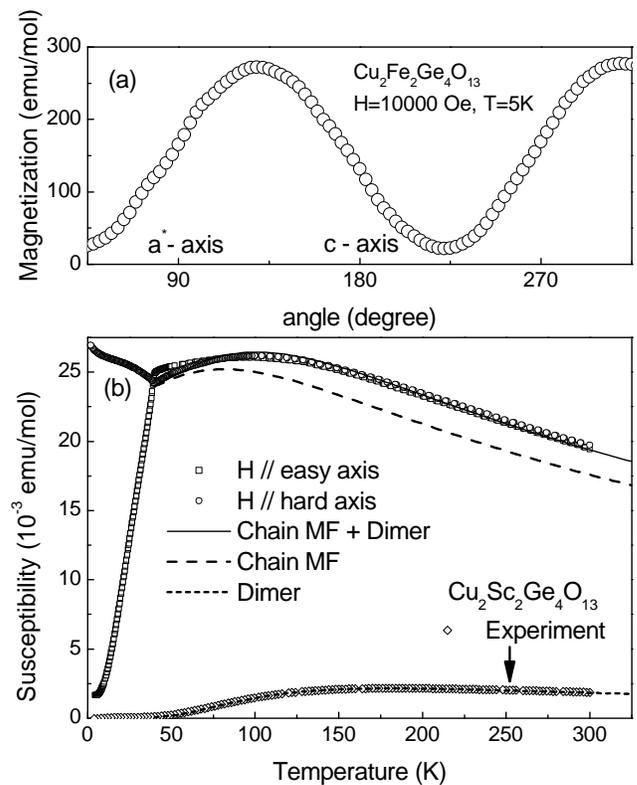}
\end{center}
\caption{(a) Angular dependence of 
magnetization in \CuFeGeO (180$^{\circ}$ 
is roughly coincident with crystallographic 
$c$ axis). (b) Magnetic susceptibility of 
\CuFeGeO\ in field oriented parallel to 
the magnetic easy and hard axes. Solid, 
dashed, and dotted lines denote theoretical 
calculations (see text). 
} \label{fig5}
\end{figure}

The magnetic susceptibility of \CuFeGeO\ 
is shown in Fig.~\ref{fig5}. 
While the behavior is qualitatively 
consistent with that seen in the 
previous study on the Fe compound,~\cite{Masuda} 
the absolute magnitude of the high-temperature 
susceptibility is approximately 15\% lower. 
As care was taken in the present study to 
ensure that there would be no magnetic 
impurities on the freshly cut bulk crystal 
surface, it is considered that the larger 
susceptibility results obtained previously 
were affected by a temperature-independent 
susceptibility component attributable to 
magnetic impurities on the surface of the 
bulk crystal. 

Prior to obtaining the present data, 
the magnetic easy axis was identified 
by measuring the angular dependence of 
the magnetization around the $b$ axis 
(Fig.~\ref{fig5}(a)). Enhanced modulation 
was observed in the $a$--$c$ plane. 
The $c$ and $a^{*}$ axes approximately 
correspond to angles of 180$^{\circ}$ 
and 90$^{\circ}$, respectively. 
The easy axis was identified by a minima 
in magnetization located approximately 
45$^{\circ}$ from the $c$ axis. This 
result is consistent with the magnetic 
structure analysis conducted previously 
for the Fe compound.~\cite{Masuda} 

The magnetic susceptibility in the easy and hard directions 
is shown in Fig.~\ref{fig5}(b). The marked decrease in susceptibility 
in the easy direction at $\lesssim 40$~K is ascribed to the N\'eel 
transition. The anisotropic behavior of typical antiferromagnets 
is clearly observed in the hard direction. 
The existence of a single anomaly 
over the entire temperature region means that the transition 
is a cooperative ordering in the bicomponent system. In the 
high-temperature region, enhancement of short-range antiferromagnetic 
correlation is indicated by the broad maximum around 100~K. 
As discussed below, the total susceptibility can be explained 
by the sum of Fe coupled chain and Cu dimer components.

\section{Analysis and Discussion}

Consistent with the crystallography, a spin dimer 
model is employed to describe the magnetic 
susceptibility of the Sc compound, as given by
\begin{equation}
\chi ^{\rm cu} = \frac{Ng^2\mu _{\rm B}^2}
{k_{\rm B}T (e^{\frac{J_{\rm Cu}}{k_{\rm B}T}} + 3)}. 
\label{dimer}
\end{equation}
Here $N$ is the number of Cu ions, $g$ is gyromagnetic ratio, 
and $J_{\rm Cu}$ is the intradimer interaction. 
The fit to the data is quite reasonable 
(dashed line, Fig.~\ref{fig2}) 
with parameters of $g = 2.02$ and 
$J_{\rm Cu} = 25.(4)$~meV.~\cite{Jdifinition} 
The dimer energy is close to that of the 
Cu mode observed in \CuFeGeO\ in the 
previous neutron scattering study.~\cite{Masuda} 
The spin framework of \CuScGeO\ is thus 
equivalent to the Cu dimers 
in \CuFeGeO. The neutron cross-section for 
spin dimers~\cite{Furrer} is given by
\begin{equation}
\frac{d^2\sigma}{d\Omega d\omega} \propto 
A(T)|f(q)|^2 (1-\frac{1}{qd} \sin qd )
\delta(\hbar \omega - J_{\rm Cu}). \label{dimerneutron}
\end{equation}
The powder average is included in the above formula. 
An intradimer distance ($d$) of $3.01$~\AA\ 
is adopted here consistent with the 
crystallography (Fig.~\ref{fig1}). 
Here, $f(q)$ is the magnetic form factor of Cu$^{2+}$ 
ions and $A(T)$ is a temperature scaling factor 
proportional to the thermal distribution of 
the singlet ground state, $1/(1+3 {\exp} (-J_{\rm Cu}/k_{\rm B}T))$. 

The dispersionless excitations observed 
in the series of constant-$q$ scans 
(Fig.~\ref{fig3}(a)) is consistent 
with the description of Eq.~(\ref{dimerneutron}), 
and the equation affords a $J_{\rm Cu}$ 
value of $24.(5)$~meV in agreement with 
the susceptibility measurements. The monotonic 
decrease and slight shoulder structure seen in 
the constant-energy scan is also reproduced 
well by this formula (Fig.~\ref{fig3}(b)). 
The temperature dependence of integrated 
intensity (Fig.~\ref{fig4}(b)) is further 
consistent with the calculation using this 
value of $J_{\rm Cu} = 24.(5)$~meV. As the 
energy width in the constant-$q$ scans 
appears to be greater than the instrumental 
resolution (Fig.~\ref{fig3}(a)), the dimers 
are weakly coupled. Interdimer interaction 
is included by replacing the delta function 
in Eq.~(\ref{dimerneutron}) with Gaussian 
and Lorentzian peak functions. The fitting 
is reasonable for both cases, affording an 
estimated intrinsic width of $3.0$--$5.0$~meV. 
This range represents the band width for Cu 
dimer excitation in \CuScGeO. 

In \CuFeGeO\, the energy scales of Fe and 
Cu spin frameworks are well separated.~\cite{Masuda2} 
The magnetic susceptibility in such a 
case can thus be described by the sum of 
individual contributions. In the lower energy 
range, the chain mean-field RPA~\cite{Scalapino} 
theory is employed for effectively coupled 
Fe spin chains. Here, Cu dimers between Fe 
chains are included in the effective interchain coupling. 
It is assumed that Fe chains on the $b$ axis are 
isotropically coupled with a coordination number $z$ of 4. 
The uniform magnetic susceptibility is then given by
\begin{equation}
\chi ^{\rm Fe} = \frac{\chi ^0}{1-zJ_{\rm Fe}'\chi ^0} 
\label{chainMF}
\end{equation}
where $J_{\rm Fe}$ and $J_{\rm Fe}'$ 
are inchain and interchain interactions 
and $\chi ^0$ 
is the susceptibility of isolated classical 
spin chains.~\cite{Fisher} In the high-energy 
range, the Cu dimer framework dominates the 
excitation. The interdimer interaction 
is small as suggested by the dispersionless 
excitation.~\cite{Masuda} It is therefore assumed 
that the susceptibility can be expressed by 
the sum of Eqs.~(\ref{chainMF}) and (\ref{dimer}). 

Fitting of the model to the Fe compound 
was performed by fixing the Cu dimer 
parameters to that of the Sc compound. 
An excellent fit to the data was thus 
obtained (Fig.~\ref{fig5}(b)), yielding 
parameters of $J_{\rm Fe} = 1.6(8)$~meV 
and $J_{\rm Fe}' = 0.17(1)$~meV. These 
values are consistent with the previous 
neutron results.~\cite{Masuda2} The data 
and associated fitting results show that 
\CuFeGeO\ is a bicomponent system, and 
that \CuScGeO\ has the same dimer component 
as the Fe compound. 

The obtained exchange parameters are 
summarized in Table~\ref{parameters}. 
The perfect coincidence between $J_{\rm Cu}$ 
of the Sc and Fe compounds shows that the 
Sc compound has the same spin dimer framework 
as the Fe compound. The remarkable difference 
between Cu modes in cooperative ordering and 
the spin liquid can also be 
observed from the respective temperature 
dependences. The peak observed for the Fe 
compound becomes substantially broader with 
increasing temperature (Fig.~\ref{fig4}(a)). 
This mode appears to be influenced by the 
magnetic ordering at 40~K. 

\section{Conclusion}

\begin{table}
 \caption{Exchange parameters for \CuScGeO\ and \CuFeGeO }
 \label{parameters} 
 \begin{ruledtabular}
 \begin{tabular}{l l l l l}
& $J_{\rm Cu}$ &  $J_{\rm Fe}$ & $J_{\rm Fe}'$ \\
 \hline
Sc (bulk) & 25.(4)~meV &  - & - \\
Sc (neutron) & 24.(5)~meV &  - & - \\
Fe (bulk) & 25.(4)~meV & 1.6(8)~meV & 0.17(1)~meV \\
Fe (neutron, Ref.~\onlinecite{Masuda}) & 24.(2)~meV & 1.60(2)~meV & 0.12(1)~meV
 \end{tabular}
 \end{ruledtabular}
 \end{table}

In summary, a combination of bulk 
measurements and neutron scattering 
experiment revealed that \CuScGeO\ is 
a mono-component compound of \CuFeGeO. 
All data were consistently explained by 
a weakly coupled dimer model. The perfect 
coincidence in the values of $J_{\rm Cu}$ 
demonstrates that \CuMGeO\ ($M$ = Fe and Sc) 
is a rare experimental realization of 
cooperative ordering and an associated 
spin liquid state. 

\begin{acknowledgments}
The authors express their great 
appreciation to R. Jin and B. C. 
Sales for heat capacity measurements, 
and A. Zheludev for fruitful discussion 
in the early stage of this study. 
The technical help of M. Matsuura, 
S. Watanabe, and K. Hirota in neutron 
scattering 
experiment is also gratefully acknowledged. 
This work was supported by a grant under 
the 2005 Strategic Research Project 
(\#W17003 and \#K17028) of Yokohama City 
University, Japan, and by a grant to 
G. J. R from the 
{\it Fond zur Foerderung der wissenschaftlichen Forschung} 
(FWF), Austria (\#R33-N10). This work was 
also partially supported by Oak Ridge National 
Laboratory (ORNL) funding under Contract 
No. DEAC05-00OR22725 (U. S. Department of Energy). 
\end{acknowledgments}


\begin{thebibliography}{14}
\bibitem{Jdifinition}
In this paper the exchange constants $J$'s are defined by 
$H = J\sum _{i} {\bm S}_{i} \cdot {\bm S}_{i+1}$.
\end{thebibliography}

\end{document}